\documentclass[aps, prd, reprint,showpacs,nofootinbib]{revtex4-1}

\usepackage{amssymb, amsmath, braket, nicefrac, graphicx} 
\usepackage[tight]{subfigure}
\usepackage{relsize}

\newcommand{\numu}{\ensuremath{\nu_\mu } }                   
\newcommand{\nue}{\ensuremath{\nu_e } }                      
\newcommand{\nutau}{\ensuremath{\nu_\tau } }                 
\newcommand{\anumu}{\ensuremath{\overline{\nu}_{\mu}} }      
\newcommand{\anutau}{\ensuremath{\overline{\nu}_\tau } }                 

\begin{document}

\title{\bf Apparent Multiple $\Delta m_{32}^2$ in 
       $\overline{\nu}_\mu$ and $\nu_\mu$ Survival Oscillations from Non-Standard
       Interaction Matter Effect}

\author{W. Anthony \surname{Mann}}
\author{Daniel     \surname{Cherdack}}
\author{Wojciech   \surname{Musial}}
\author{Tomas      \surname{Kafka}}
\affiliation{Tufts University, Medford, MA 02155}

\pacs{14.60.Lm, 14.60.Pq, 14.60.St, 96.40.Tv}

\begin{abstract}
Neutrinos propagating through matter may participate in forward coherent neutral-current-like scattering arising from non-standard interactions as well as from the Mikheyev-Smirnov-Wolfenstein matter potential $V_e$. 
We show that at fixed long baselines through matter of constant density,
the non-standard interaction potential $\epsilon_{\mu\tau} V_e$ can contribute an additional term to the oscillation phase whose sign differs for $\anumu$ versus $\numu$ propagation in matter.  
Its presence can cause different apparent $\Delta m^2$ to be erroneously inferred on the basis of oscillations in vacuum, with values lying above (for $\anumu$) or below (for $\numu$) the actual $\Delta m^2_{32}$ for the case where  $\epsilon_{\mu\tau}$ is
predominantly real-valued and of sign opposite to $\Delta m_{32}^2$.
An NSI scenario invoking only $\Re(\epsilon_{\mu\tau})$ is shown to be capable of accounting for  
a disparity recently reported between oscillation survival for $\anumu$ and $\numu$ fluxes measured at $735~\mathrm{km}$ by the MINOS experiment.  Implications for mantle traversal by atmospheric neutrinos are examined. 
The NSI matter potential with non-maximal mixing could evade conventional atmospheric neutrino analyses which do not distinguish $\numu$ from $\anumu$ on an event-by-event basis. 
\end{abstract}

\maketitle

\section{Introduction} \vspace{-9pt}
Accelerator-based neutrino long baseline experiments have recently entered an era wherein oscillation measurements 
using antineutrino exposures are being undertaken.   Since previous experimental investigations have been have based in the
main on neutrino beam exposures, the on-going and next-round explorations using antineutrino reactions will provide
a crucial complementary perspective on oscillation phenomena.   The question naturally arises as to whether standard phenomenology, describing mixing among either two or three of active flavor neutrinos and without or with inclusion of the
Mikheyev-Smirnov-Wolfenstein (MSW) matter effect \cite{MSW}, provides an adequate framework -- or whether a more expansive formalism is required.     In this work we investigate a matter effect scenario which is based upon a specific neutrino 
non-standard interaction (NSI) which we identify as highly promising for generating distinctive observable effects in the 
neutrino ``atmospheric" $\mu$-$\tau$ flavor mixing sector.   

Impetus for this paper arises in part from observations of muon-neutrino flavor disappearance for $\anumu \rightarrow \anumu$ and for $\numu \rightarrow \numu$ obtained by the MINOS experiment.   As recently reported  \cite{Tricia},  MINOS infers the value 
for $\Delta m_{32}^2$ governing $\anumu$ oscillations at $735~\mathrm{km}$
to be somewhat higher than the value gleaned from $\numu$ disappearance.    The
new observations are
in line with the MINOS low statistics result of last year, in which $\anumu \rightarrow \anumu$ survival was examined using 
antineutrino (background) reactions isolated in exposures with neutrino beams~\cite{Hartnell}.   
While the earlier observation was of poor statistical significance, it attracted attention as a possible harbinger of new physics, e.g. of CPT violation in the neutrino sector~\cite{Barenboim}, or of new interactions with matter-driven mixing to a sterile neutrino~\cite{Nelson}, or of an NSI matter effect~\cite{Mann-Cherdack}.

We begin with a brief summary in Sec. II of salient points from the literature on NSI matter effects 
as they pertain to neutrino oscillations.  In Sec. III we then specify
the neutrino oscillation phenomenology which follows from existence of a single NSI matter potential $\epsilon_{\mu\tau} V_e$, proposed to exist in addition to the MSW matter effect.  
The latter, well-known matter potential is $V_{e} = \sqrt{2} G_F n_{e}$, where $G_F$ is the Fermi constant, and $n_{e}$ is the electron number density averaged over the neutrino path through the Earth.   
This sets the stage for discussion of certain remarkable implications for experiments in Sec. IV.   Our formalism applies most directly to 
neutrino traversal of the Earth's crust, and  we use recent  results  from the MINOS long baseline experiment to 
illustrate such an application.
We proceed to show that a real-valued $\epsilon_{\mu\tau}$ in conjunction with a single $\Delta m^2_{32}$ and with a single mixing angle $\theta_{23}$ provides a satisfactory description for the MINOS antineutrino and neutrino beam data currently available.
In Sec. V we relate our formalism to
neutrino traversal of the Earth's mantle, and identify effects of relevance to measurements using atmospheric
neutrinos.    Section VI concludes the paper with mention of near-future experimental developments which can  
shed light on the existence of $\epsilon_{\mu\tau}$, and on other new physics mechanisms  
which also have capability to differentiate $\anumu$ from $\numu$ oscillations.

\vspace{-9pt}
\section{Neutrino NSI matter effect} \vspace{-9pt}
Under the hypothesis that neutrinos participate in heretofore unobserved non-standard interactions, the neutrino flavor Hamiltonian will carry new potential terms analogous to the MSW potential.
The latter potential accounts for coherent forward scattering of electron-flavor neutrinos from the electrons within a field of matter.
Possibilities for the underlying neutrino scattering processes include both flavor-changing and flavor-conserving NSI.
There are six possible NSI amplitudes which can arise in neutrino propagation through matter, conventionally designated as $\epsilon_{ee}$, $\epsilon_{e\mu}$, $\epsilon_{e\tau}$, $\epsilon_{\mu\mu}$, $\epsilon_{\mu\tau}$, and $\epsilon_{\tau\tau}$. 
CP-violating effects are possible in the event that the amplitudes carry phases. 
The effects which may arise from various combinations of $\epsilon_{\alpha\beta}$ have received extensive treatment in the literature;
recent summaries with relevant references can be found in Refs.~\cite{Minakata, Yasuda}.

For oscillation survival probabilities of the $(\numu, \nutau)$ sector of interest here, direct contributions may arise from the three potentials ~$\epsilon_{\mu \mu}V_e$, ~$\epsilon_{\tau \tau}V_e$,  and  ~$\epsilon_{\mu \tau}V_e$.  
In a recent evaluation of model-independent bounds on production and detection of neutrino NSI it was found that $|\epsilon_{\mu\mu}| < 0.064$,  whereas $|\epsilon_{\mu\tau}| < 0.33$ and $\epsilon_{\tau\tau}$ is even less constrained~\cite{Biggio}.   
We note that limits on $|\epsilon_{\tau\tau}^{eL}|$ and $|\epsilon_{\tau\tau}^{eR}|$ based on solar and KamLAND neutrino data
are reported from an analysis which restricts to neutrino NSI with electrons (but not with u or d quarks)~\cite{Bolanos}.

In this work we neglect the flavor-conserving NSI and focus on $\epsilon_{\mu \tau}V_{e}$ as the single NSI capable of producing significant differences in $\anumu$ versus $\numu$ oscillations in long-baseline experiments.   The extraordinary role that
$\epsilon_{\mu\tau}$ can play in distinguishing $\anumu$ from $\numu$ oscillations has received little direct discussion in 
the literature to date.  This circumstance reflects in part a dearth of sufficient experimental information to motivate NSI
expositions to get down to specifics.   However a recent work has discussed a somewhat analogous role for the
$\epsilon_{e\tau}$ NSI (of magnitude $\sim$10$\%$) with explaining discordant $\theta_{12}$ values inferred from solar (electron) neutrinos versus KamLAND (electron) antineutrinos~\cite{Palazzo}.

A two-flavor neutrino mixing framework is an adequate venue for
examination of $\anumu$ and $\numu$ propagation in the constant-density terrestrial crust, 
and we invoke such a framework in Sec. III.    It has been argued, on the basis of fitting using the two-flavor mixing 
framework to SuperKamiokande and MACRO atmospheric ($\numu$ + $\anumu$) data, that the trend of muon-flavor survival at high energies and for global baselines constrains $|\epsilon_{\mu\tau}|$ to be a few percent or less \cite{Fornengo, Garcia-Maltoni}.
For neutrino and antineutrino propagation over globe-spanning baselines however, there are complications.
Conventional three-flavor oscillations requires treatment of the different
matter densities of the crust, mantle, and core, with the MSW resonances of the mantle and core coming into play.   With inclusion of
NSI matter effects into the ``mix",  e.g. $\epsilon_{e\tau}$ as well as $\epsilon_{\mu \tau}$,  significant new degrees of 
freedom become available which can couple to ones neglected in a two-flavor mixing framework.   It has been demonstrated for
$\epsilon_{e\tau}$,  $\epsilon_{\tau\tau}$, and $\epsilon_{ee}$, that NSI bounds derived from atmospheric neutrinos via two-flavor analysis become relaxed when $\nue$ mixing is included~\cite{ FLM04, FL05}.    The possibility also remains that a similar
outcome would ensue in a full three-flavor analysis extending to $\epsilon_{\mu\tau}$.   Consequently, the verity 
of stringent $\epsilon_{\mu\tau}$ bounds obtained using the two-flavor mixing framework for atmospheric neutrinos has been
called into question~\cite{Biggio, OPERA}.   A three-flavor framework treatment for atmospheric neutrinos with NSI is generally regarded to be preferable~\cite{Akhmedov}.  
In the absence of a comprehensive three-flavor framework analysis at this time, 
we take the view that $|\epsilon_{\mu\tau}|$ values exceeding the two-flavor mixing limits, while possibly disfavored, 
are not as yet ruled out. 

As we elaborate in Sec. V, the phenomenology  allows structure which only becomes apparent when $\anumu$ samples are treated separately from $\numu$ samples~\cite{OPERA}.   More generally,  our study suggests
that, in fitting an NSI $\epsilon_{\mu\tau}$  scenario to data for which no distinction between $\anumu$ and $\numu$
on an event-by-event basis is available, there is a risk  
of ``averaging'' over NSI structure 
in such way as to reduce to a mimicry of conventional vacuum oscillations.

\vspace{-10pt}
\section{Two-flavor evolution with matter effect} \vspace{-9pt}
To obtain accessible expressions of sufficient accuracy, we neglect solar-scale mixing ($\Delta m^{2}_{21} = 0$) and subdominant $\numu \rightarrow \nue$ oscillations ($\theta_{13} = 0$), and work in a two-state mixing framework.  
Unless otherwise noted, the normal hierarchy for neutrino mass eigenstates is assumed.  
Basis states for neutrino flavor ($\alpha = \mu , \tau$) are then $\{  \ket{\numu}, \ket{\nutau} \}$  and the evolution of states in time ($\hbar = c = 1$, so $t = L$ for ultrarelativistic neutrinos) is governed by the effective wave equation 
\begin{equation} \label{eq:schroedinger_matter}
i \frac{d}{dt} \vec \nu^{(\alpha)}(t) = \hat H^{(\alpha)} \vec \nu^{(\alpha)}(t) .
\end{equation}

The Hamiltonian in flavor basis describing neutrino propagation in vacuum is obtained from the vacuum Hamiltonian in mass basis $\hat H^{(23)}_{0}=\text{diag}(0, \Delta m_{32}^2/2E_\nu)$ through a rotation via the standard unitary mixing matrix $\hat R_1(\theta_{23})$. 
We augment the flavor basis vacuum Hamiltonian with the NSI potential term $\hat H^{(\alpha)}_{\textrm{matter}}$
\begin{equation} \label{eq:hamiltonian_matter}
  \hat H^{(\alpha)}_{\textrm{matter}}
    = \left( \begin{array}{cc}
          0 & \epsilon_{\mu \tau}V_{e} \\
          \epsilon_{\mu \tau}^{*}V_{e} & 0 \\
      \end{array} \right) .
\end{equation}

\noindent The full Hamiltonian is then given by
\begin{equation}
\hat H^{(\alpha)} = \hat R_1 \cdot \hat H_0^{(23)} \cdot \hat R_1^\mathtt{T} + \hat H_{matter}^{(\alpha)}
\end{equation}

Only the real part of $\epsilon_{\mu\tau}$ distinguishes between $\numu$ and $\anumu$ in the derivations to follow. 
Hereafter we neglect the CP-violating imaginary part of $\epsilon_{\mu\tau}$ and assume $\epsilon_{\mu\tau}=\Re(\epsilon_{\mu\tau})$ in order to focus on the physics implied by the real part of $\epsilon_{\mu\tau}$.

After algebraic manipulation leading to removal of a term proportional to $\mathbb{\hat I}$, which merely contributes an overall phase to the oscillation amplitudes, the Hamiltonian can be expressed as
\begin{equation}
\hat H^{(\alpha)} = \vec N \cdot \vec \sigma
\end{equation}

\noindent where $\vec \sigma$ is the Pauli vector, and 
\begin{equation}
\vec N = \left(  \sin(2\theta_{23}) \frac{\Delta m^2_{32}}{4 E_\nu} +  \epsilon_{\mu \tau} V_{e}, 0,  -\cos (2\theta_{23})\frac{\Delta m^2_{32}}{4 E_\nu}      \right) .
\end{equation} 

The corresponding evolution operator $\hat U(t=L,0)$ is
\begin{equation} \label{eq: evolution op}
\hat U(t, 0) ~= e^{-i \hat H^{(\alpha)}t } ~= e^{-i \vec N \cdot \vec \sigma \; L} = e^{-i \vec n \cdot \vec \sigma \; \phi}   .
\end{equation}

\noindent
The evolution operator $\hat U$ amounts to a rotation of $2\phi$ about the direction $\vec n \equiv \vec N / |\vec N|$ in the two-flavor spinor space, where $\phi \equiv |\vec N| L$ is given by
\begin{eqnarray}  
   \phi = \Bigg\{ \left(\frac{\Delta m^{2}_{32}}{4 E_{\nu}}\right)^{2} 
    \mp 2 \sin (2\theta_{23}) \; \frac{\Delta m^{2}_{32}}{4 E_{\nu}} \;  \epsilon_{\mu \tau}|V_{e}|   \nonumber \\  
    \label{eq:oscillation-phase} 
    + \big( \epsilon_{\mu \tau} V_{e} \big)^{2} \Bigg\}^{1/2} L.
\end{eqnarray}

In Eq. (\ref{eq:oscillation-phase})  the middle term admits both signs, depending on whether it describes $\numu$ or $\anumu$. 
This is because the matter potential $V_e$ becomes negative for $\anumu$ propagation.
(In Eq. (\ref{eq:oscillation-phase}), and in Eqs. (\ref{eq:oscillation-intensity}), (\ref{eq:bottom-line}) to follow, the upper (lower) sign refers to $\anumu$ ($\numu$) oscillations, whereas $\epsilon_{\mu \tau}$ carries its own fixed sign.)

From the time evolution operator, the survival probability of $\numu$ ($\anumu$) at $t=L$ is
\begin{equation} \label{eq:survival-probability-2}  
  \mathcal{P}\left(\,^{(} \overline \nu_\mu ^{)} ~\rightarrow ~^{(}\overline \nu_\mu ^{)}\,\right)
    ~= 1 ~- \mathcal{F}\sin^2\phi ,
\end{equation}  

\noindent          
where
\begin{equation} \label{eq:oscillation-intensity}  
  \mathcal{F}= 
     1- \frac{\cos^{2}2\theta_{23}}{ 
    1 ~\overline + ~ 2\sin(2\theta_{23})  \frac{4 \epsilon_{\mu \tau}|V_{e}|}{\Delta m^{2}_{32}} E_{\nu}    
  + \left( \frac{4 \epsilon_{\mu \tau}V_{e}}{\Delta m^{2}_{32}} E_{\nu} \right)^{2}    } .
\end{equation}  

It can be seen for the case $\epsilon_{\mu\tau}=0$ that $\mathcal{F}$ reduces to $\sin^2 2\theta_{23}$, and $\phi$ reduces to the vacuum phase, hence the vacuum survival probability is recovered.                  
For the case of maximal mixing, $\theta_{23} = 45^{0}$ and $\cos 2\theta_{23} = 0$, the $\mathcal{F}$-factor becomes 1 and the survival probabilities simplify to {\setlength\arraycolsep{0.0em}
\begin{eqnarray} \label{eq:bottom-line} 
  \mathcal{P}\Big(\,&&^{(} \overline \nu_\mu ^{)} \rightarrow^{(}\overline \nu_\mu ^{)}\,\Big) \simeq \nonumber              \\ 
    &&1 - \sin^2  \left( \left| ~\frac{\Delta m^{2}_{32}}{4 E_{\nu}} 
    ~\overline +  ~\epsilon_{\mu \tau} |V_{e}| ~\right |  L \right).
\end{eqnarray}
}
  
\noindent    
We observe that relative to the standard survival probability, the vacuum oscillation phase is augmented by an additional term which is independent of $E_{\nu}$;
it is a matter effect induced by the flavor-changing NSI.   
It is important that the NSI matter effect phase term appears with a different sign in the expressions for $\numu$ and $\anumu$ survival probabilities. 

Our formalism as above is Rabi spin oscillations in another guise;   it utilizes 
NSI phenomenology which is contained implicitly in many published expositions.   However we
do not find another treatment which explicitly develops the phenomenology of
$\epsilon_{\mu\tau}$ to expressions which are as convenient as
Eqs. (\ref{eq:oscillation-phase}), (\ref{eq:survival-probability-2}),  and (\ref{eq:oscillation-intensity}).

\subfigtopskip=3pt
\subfigbottomskip=0pt
\begin{figure*}[!htb]
\begin{center}
\subfigure{\includegraphics[width=16cm]{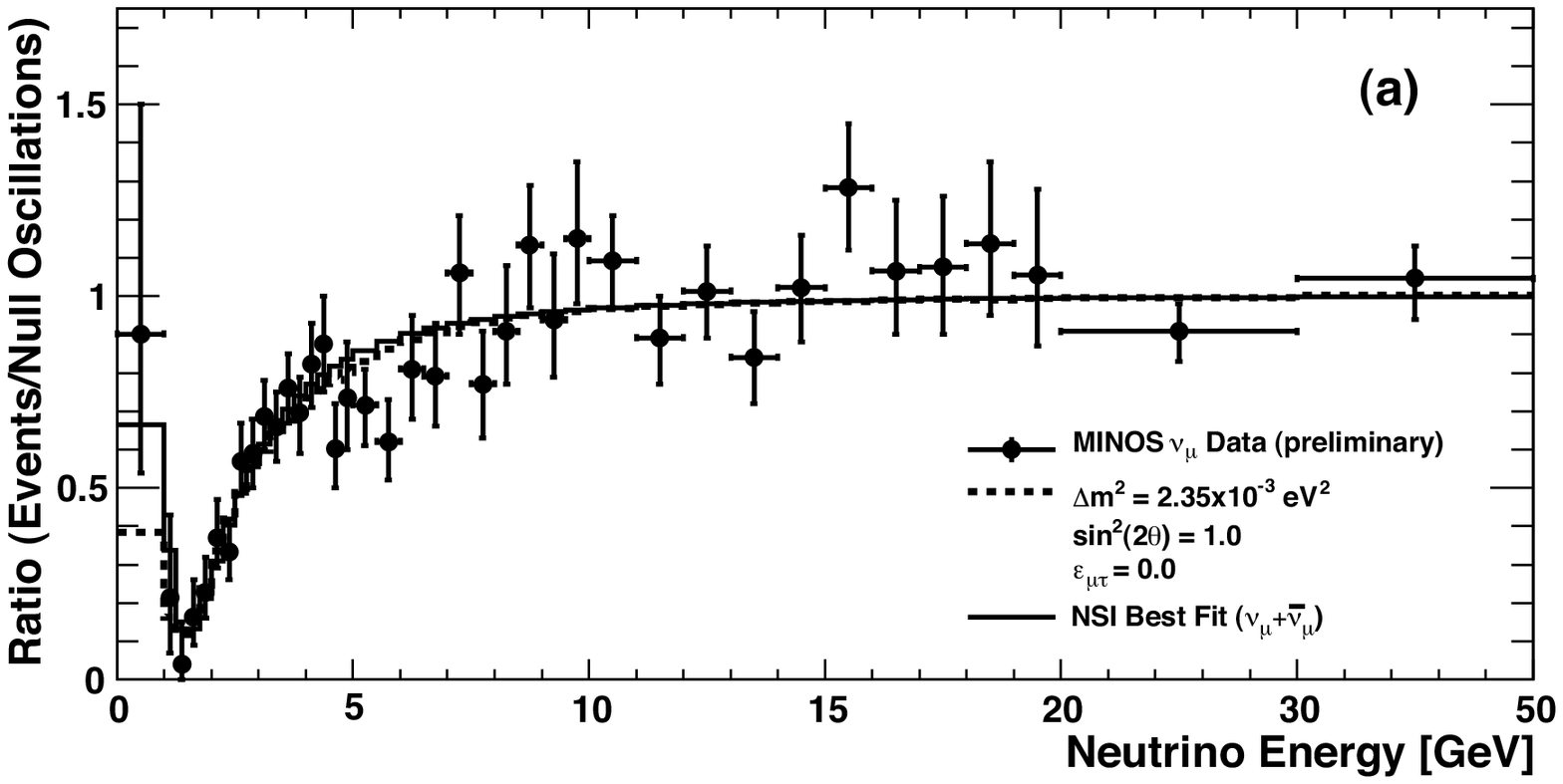} \label{fig:surv1}}
\subfigure{\includegraphics[width=16cm]{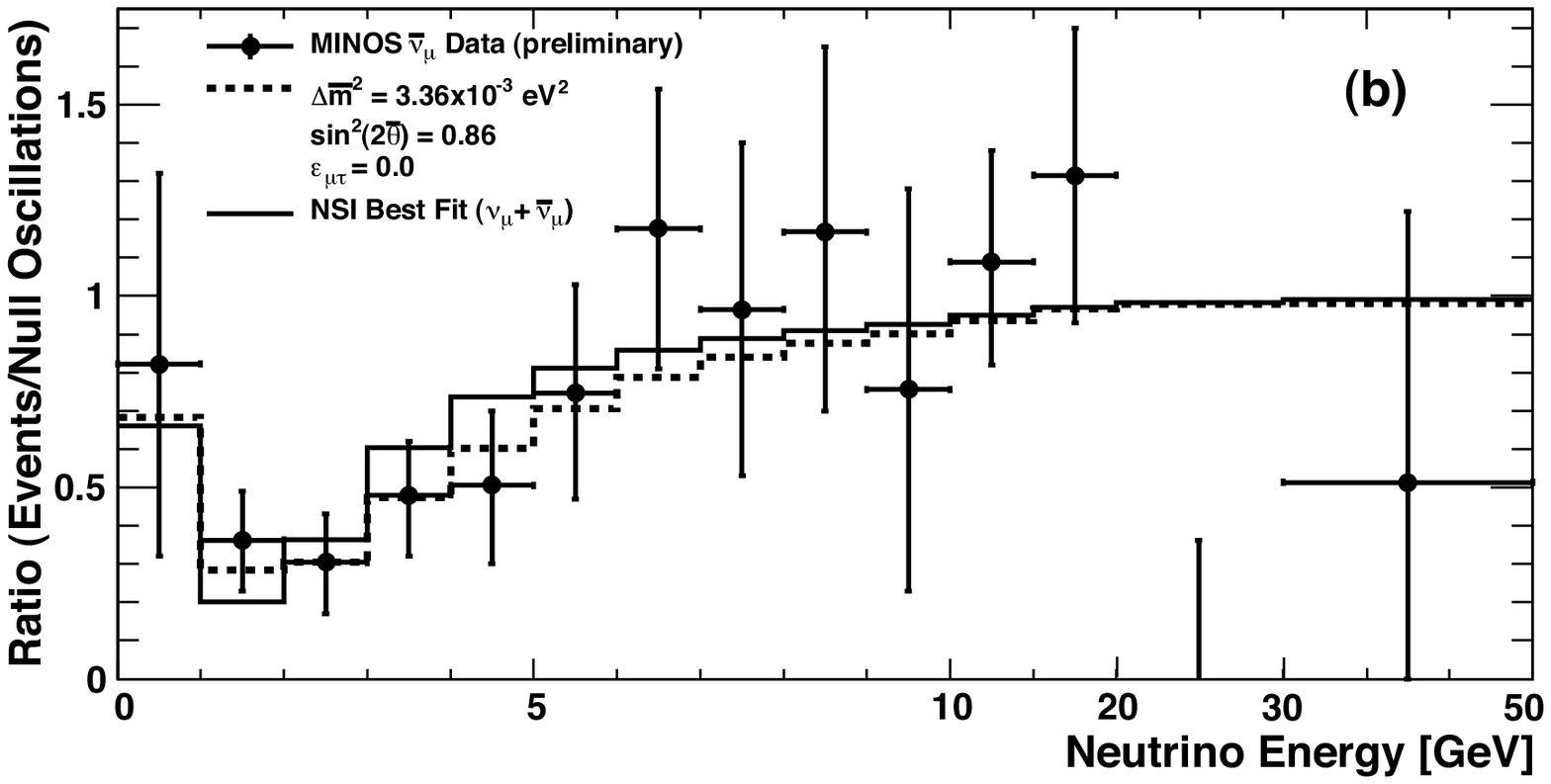} \label{fig:surv2}}
\end{center}
\caption{\small 
Histogram 1(a) $\big(\mathrm{1(b)}\big)$ shows the survival probability averaged per bin for $\numu \rightarrow \numu$ ($\anumu \rightarrow \anumu$) at 735 km. 
The MINOS data points (Ref.~\cite{Tricia}) show the ratios of observed CC-$\numu$ (CC-$\anumu$) event rates to rates predicted by MINOS in the absence of oscillations. 
In each Figure, the solid histogram shows the result of fitting the matter effect $\numu$ ($\anumu$) survival probability of Eq. (\ref{eq:survival-probability-2}) to the MINOS $\numu$ and $\anumu$ oscillation survival ratios. 
The dashed histogram shows the vacuum oscillations $\numu$ ($\anumu$) survival probability for parameters of each of the 
MINOS fits (carried out separately for each sample).}
\vspace{-0.5cm} 
\end{figure*}

\begin{figure}[!htb]
\begin{center}
\includegraphics[width=9.6cm]{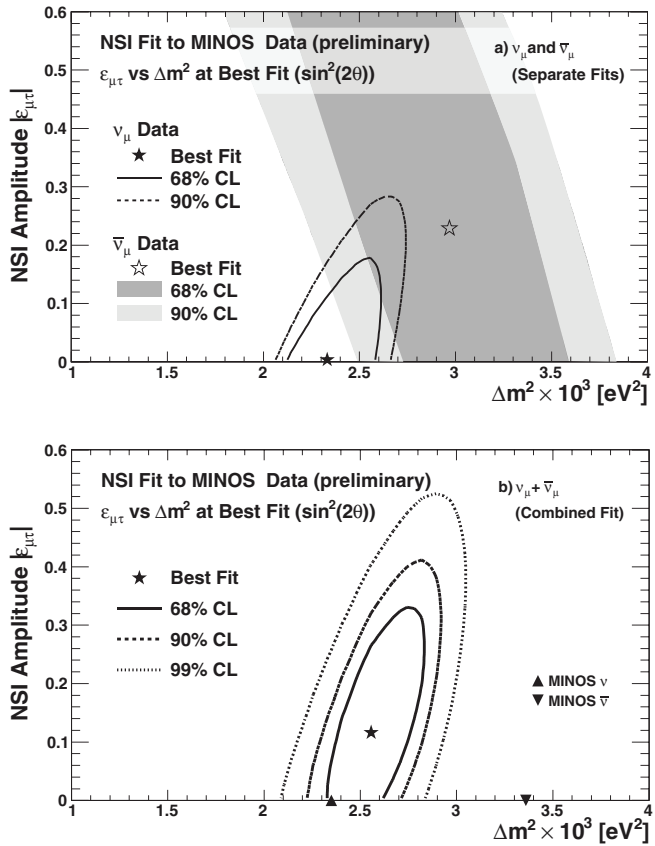}
\caption{
\small Contours in $\Delta \chi^{2}$ for $|\epsilon_{\mu \tau}|$ versus $\Delta m^{2}_{32}$ from fits to the $\anumu$-CC and $\numu$-CC survival probability ratios as functions of $E_{\nu}$ observed at the MINOS far detector (Ref.~\cite{Tricia}). 
Plot (a) shows the contours from separate fits to $\numu$ (ellipsoidal regions) and $\anumu$ (shaded regions).   
Plot (b) shows the three-parameter combined fit to both $\numu$ and $\anumu$.  Here, the correlation properties exhibited by 
the separate fits of Plot (a) tend to merge for non-zero, negative values of $\epsilon_{\mu \tau}$.    In both plots the best fit points are denoted by stars;  the up- (down-) pointing triangle in Plot (b) depicts the vacuum oscillation result of Ref.~\cite{Tricia} for $\numu$ ($\anumu$).
}
\label{fig:NSI-ME-fit}
\end{center}
\vspace{-0.4cm}
\end{figure}

\begin{figure}[!htb]
\begin{center}
\includegraphics[width=8.5cm]{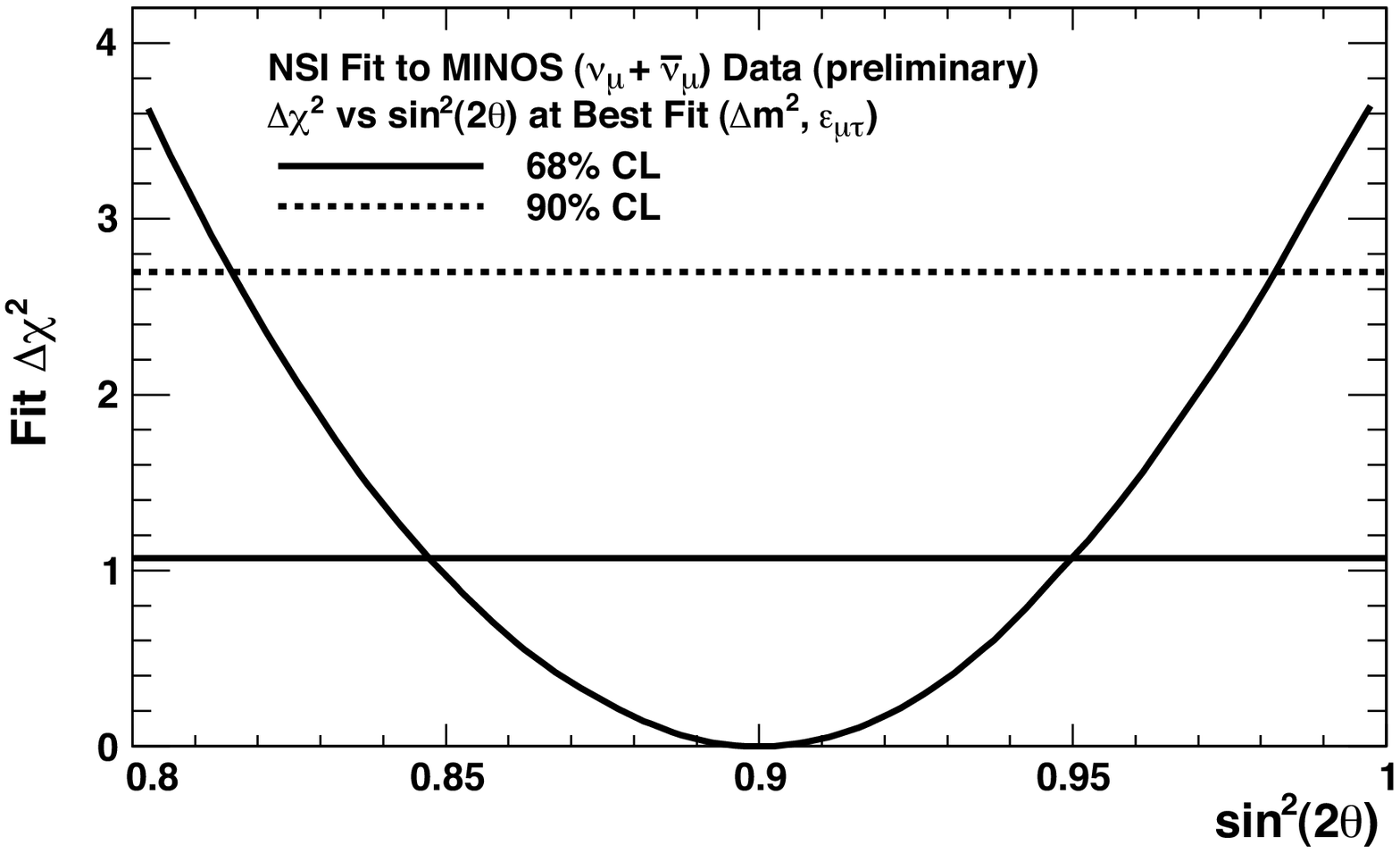}
\caption{
\small The $\Delta \chi^{2}$ for $\sin^{2} 2\theta_{23}$ with $\Delta m^{2}$ and $\epsilon_{\mu\tau}$ marginalized.   
The fit is to the MINOS data ratios versus $E_{\nu}$ which expresses $\numu$ and $\anumu$ survival probabilities at $735~\mathrm{km}$ (Ref.~\cite{Tricia}).   
The  solid (dashed) horizontal line shows the contour boundaries at the $68\%$ ($90\%$) confidence level.}
\label{fig:mixing-angle}
\end{center}
\vspace{-0.7cm}
\end{figure}

 \vspace{-10pt}
\section{Implications for Crust Traversal}
\vspace{-9pt}
\subsection{MINOS baseline}
The recent MINOS results~\cite{Tricia} are based on NuMI neutrino beam exposures of $ 1.7 \times 10^{20}$ protons-on-target (PoT) and $ 7.2 \times 10^{20}$ PoT using Reverse and Forward Horn Current running (hereafter referred to as the RHC and FHC exposures).
It is reported that the mass-squared difference of neutrino mass states, $\Delta \overline{m}^2$, deduced from the observed $\anumu$ survival rate at the MINOS far detector at $735~\mathrm{km}$ has best fit value of $ 3.36^{+0.45}_{-0.40} \times 10^{-3}~\mathrm{eV}^{2}$. 
This value is higher than the updated MINOS best fit value for $\numu$ disappearance, $\Delta m^2 =  2.35^{+0.11}_{-0.08} \times 10^{-3}~\mathrm{eV}^{2}$.
Both $\anumu$ and $\numu$ data are compatible with maximal mixing, however the best fit for $\anumu$ allows more deviation: $\sin^{2} 2\overline{\theta} = 0.86 \pm 0.11$ for $\anumu$ compared to $ \sin^2 2\theta > 0.91$ (90\% CL) for $\numu$ survival.

The apparent disparity between $\numu$ and $\anumu$ data can be accounted for with the NSI matter effect.
As seen from Eq. (\ref{eq:oscillation-phase}), the oscillation phase receives a matter-induced contribution which retards (advances) the total phase for propagating $\numu$ ($\anumu$). 
This shift in oscillation phase causes the observed survival probability to mimic vacuum oscillations with a $\Delta m^2$ of lower (higher) value. 
Consequently, the actual value of $\Delta m_{32}^2$ according to our NSI formalism lies above (below) the best fit value of Ref. \cite{Tricia}. 
In the case of normal mass hierarchy, the NSI potential $\epsilon_{\mu \tau}V_e$ needs to be negative (positive) for propagating $\numu$ ($\anumu$) and so $\epsilon_{\mu\tau}$ must be negatively-valued. On the other hand, for inverted hierarchy with $\Delta m^2_{32} < 0$, $\epsilon_{\mu\tau}$ would need to have a positive value. 
For the MINOS $735~\mathrm{km}$ baseline through the Earth's crust of density $\rho \simeq 2.72\; \mathrm{g/cm}^{3}$, we use $V_{e} \simeq 1.1 \times 10^{-13}\;\mathrm{eV} = (1/1900)~\mathrm{km}^{-1}$~\cite{OPERA}.

For purposes of fitting, the slides of Ref.~\cite{Tricia} were magnified to extract the data points and the error ranges presented.   
For instances of asymmetric errors, the larger of the two is chosen.  
A $\chi^2$ fit is performed over the MINOS data ratios which express the oscillation probabilities.  
The data is fitted to the phenomenological predictions of Eqs. (\ref{eq:oscillation-phase}), (\ref{eq:survival-probability-2}), and (\ref{eq:oscillation-intensity}) with $\Delta m_{32}^2$, $\sin^2 2\theta_{23}$, and $\epsilon_{\mu \tau}$ treated as free parameters.
The role of these parameters is defined by the NSI phenomenology, which reduces to vacuum $\nu_{\mu} \rightarrow \nu_{\tau}$ mixing as $|\epsilon_{\mu \tau}|$ approaches zero.

Figure~\ref{fig:surv1} shows the MINOS neutrino data as ratios of the observed rate of events at $735~\mathrm{km}$ to the Monte Carlo expectation in the absence of oscillations.
The binned ratios are shown as a function of event reconstructed energy $E_{\nu}$ for $\numu$ charged current (CC) events recorded in the MINOS far detector from FHC exposures totaling $7.2 \times 10^{20}\; \mathrm{PoT}$.   
Superimposed (solid line) is the binned matter effect survival probability for the best-fit parameters as determined by both $\numu$ and $\anumu$ data (described below), using Eq. (\ref{eq:survival-probability-2}).
We find: 
$\big(\,\Delta m^{2}_{32},\, \sin^{2} 2\theta_{23},\, \Re(\epsilon_{\mu\tau})\,|V_{e}|\big) = \big(\,2.56^{+0.27}_{-0.24} \times 10^{-3}~\mathrm{eV}^{2},\, 0.90 \pm 0.05,\, -(0.12\pm 0.21)\,|V_{e}|\big)$, where $V_e$ is evaluated for terrestrial crust. 
Also shown for comparison (dashed line) is the fit to $\numu \rightarrow \numu$ vacuum oscillations obtained by MINOS. The best fit parameters are $(2.35 \times 10^{-3}~\mathrm{eV}^{2}, 1.0, 0.0)$. 

Fig.~\ref{fig:surv2} shows the corresponding distribution of ratios for antineutrino events observed over the null oscillation expectation, as a function of $E_{\nu}$ for $\anumu$ charged current events of the MINOS RHC exposure of $1.7\times 10^{20}\; \mathrm{PoT}$.
The solid curve displays the best fit as shown in Fig.~\ref{fig:surv1}, differing only in the sign of $\epsilon_{\mu\tau}\, |V_e|$.  
Shown for comparison (dashed line) is the MINOS result obtained by fitting with 
$\anumu \rightarrow \anutau$
vacuum oscillations; the best fit oscillation parameters are $(3.36 \times 10^{-3}\;\mathrm{eV}^2, 0.86, 0.0)$.  

Figures~\ref{fig:NSI-ME-fit}(a), \ref{fig:NSI-ME-fit}(b) show  confidence level (CL) contours for $|\epsilon_{\mu\tau}|$ versus $\Delta m^{2}$ from fitting to the MINOS samples with $\sin^2 2 \theta$ marginalized (at each point, the non-displayed fit parameters are chosen so as to yield the lowest $\chi^2$).  In each plot,
contours derived from the various fits are defined as the set of points on the $|\epsilon_{\mu \tau}|$ versus $\Delta m^2$ plane where $\Delta \chi^2$, the difference between $\chi^2$ and the best fit $\chi^2$, is $2.3$ ($68\%$ CL), $4.61$ ($90\%$ CL), or $9.21$ ($99\%$ CL).   Figure~\ref{fig:NSI-ME-fit}(a) shows $|\epsilon_{\mu\tau}|$ versus $\Delta m^{2}$ contours which are determined separately for the neutrino and antineutrino samples.    Fitting to the $\numu$ data yields the ellipsoidal contours which surround the best-fit point
(solid star) at $(2.34\times 10^{-3} \mathrm{eV}^2, 0.94, -0.004|V_{e}|)$.    The contours exhibit a positive correlation between higher
$\Delta m^2$ values and larger (more negative) $|\epsilon_{\mu\tau}|$.  Also shown are the corresponding contours obtained from 
fitting the $\anumu$ data, which surround the best-fit point at $(2.98\times 10^{-3} \mathrm{eV}^2, 0.83, -0.23|V_{e}|)$.  
In contrast to the correlation property of the $\numu$ contours, the $\anumu$ contours correlate lower $\Delta m^2$ values with
larger $|\epsilon_{\mu\tau}|$.   
The $\anumu$ contours are also much broader - less restrictive - than those from the $\numu$ fit, providing a striking visual reminder that $\anumu$ event statistics are relatively poor compared to $\numu$ statistics in currently available long baseline exposures.
Nevertheless one can see from Fig.~\ref{fig:NSI-ME-fit}(a) that the opposite-leaning correlations exhibited by the two independent
sets of contours will tend to merge for non-zero, negative values of $\epsilon_{\mu\tau}$ when both $\numu$ and $\anumu$ are
fitted simultaneously.

Figure~\ref{fig:NSI-ME-fit}(b) displays the $\Delta \chi^2$ contours and the best fit location on the $|\epsilon_{\mu\tau}|$ versus $\Delta m^{2}$ plane from the joint fit to the $\numu$ and $\anumu$ data.
The fit converges to an `intermediate' $\Delta m_{32}^2$ value with the NSI matter coupling negatively-valued and of 12\% of the strength of the conventional MSW matter effect.   Figure~\ref{fig:mixing-angle} shows the $\Delta \chi^{2}$ obtained from fitting to $\sin^{2} 2\theta_{23}$ with the parameters of Figure~\ref{fig:NSI-ME-fit}(b) marginalized.  
Non-maximal mixing, namely $\sin^{2} 2\theta_{23} = 0.90$, is preferred;  $\sin^{2} 2\theta_{23} = 1.0$ is excluded at greater than $90\%$ CL.

The $\chi^2$ per degrees of freedom for the three-parameter fit to both samples is $44.1/50 = 0.88$. The fitted data comprises bin distributions of 1986 $\numu$ events together with only $97$ of $\anumu$ events. Obviously, the MINOS $\numu$ sample - at present - exerts dominant statistical power within the fit.    Our three-parameter fit does not constitute supporting
evidence for a claim concerning the $\epsilon_{\mu \tau}$ NSI.  On the contrary,
it is readily discerned in Fig.~\ref{fig:NSI-ME-fit}(b) 
that solutions with $\epsilon_{\mu \tau} = 0$ lie within the one-sigma contour, 
hence the NSI fit provides a description which is no better (or worse) that that afforded  
by conventional vacuum oscillations.   Rather, the NSI fit to MINOS data serves to
illustrate how the phenomenology may play out upon inclusion of larger $\anumu$ oscillation samples.

The dashed histograms of Figs.~\ref{fig:surv1} and \ref{fig:surv2} show the descriptions afforded by the separate vacuum oscillation solutions obtained by the MINOS collaboration.  Our NSI scenario 
provides a characterization of both data sets using only three oscillation parameters whose relationship to each set is specified by the phenomenology proposed.   
This is in contrast to the vacuum oscillation descriptions using two different pairs of parameters,  namely $\Delta \overline{m}^{2}$, $\sin^{2} 2 \overline{\theta}$ and $\Delta m^{2}$, $\sin^{2} 2\theta$.

\subsection{K2K and T2K baselines}
In view of the $\epsilon_{\mu\tau}$ matter effect scenario presented here, it is sensible that the K2K experiment reported a value of $\Delta m^{2}$ higher than that obtained with MINOS ~\cite{Tricia, MINOS-CC}. That is, the K2K best fit value is $2.8 \times 10^{-3}\;\mathrm{eV}^{2}$ with $\sin^2 2\theta = 1.0$; at $90\%$ CL, the allowed $\Delta m^2$ range is between $1.9$ and $3.8 \times  10^{-3}\; \mathrm{eV}^2$ with $\sin^2 2 \theta > 0.6$~\cite{K2K}.   
This is because the K2K baseline of $250\;\mathrm{km}$ involves less matter traversal than does neutrino propagation in MINOS, and also the neutrino beam spectrum is confined to lower energies,  $0.5 < E_{\nu} < 3~\mathrm{GeV}$.  
The result is that the vacuum part of the oscillation phase will dominate the perturbing term from the matter effect.   
In the future, when the T2K experiment with $295~\mathrm{km}$ baseline obtains a sizable $\numu$ exposure, it will be of interest to include those $\numu \rightarrow \numu$ results into fitting with the NSI $\epsilon_{\mu\tau}$ scenario.

\vspace{-10pt}
\section{Implications for mantle traversal}
\vspace{-9pt}
\subsection{Phenomenology}
The matter effect survival probabilities proposed for $\numu \rightarrow \numu$ and for $\anumu \rightarrow \anumu$ have implications for longer neutrino propagation baselines, such as those of atmospheric neutrinos which may travel thousands of kilometers through the terrestrial mantle and core.
Given the elevated densities of the matter fields traversed, one infers from Eq. (\ref{eq:oscillation-phase}) that apparent differences in the values of $\Delta m^2$ for $\numu$ versus $\anumu$ oscillations may be more pronounced than for neutrino baselines confined to the terrestrial crust.   
The effects described by 
Eqs. (\ref{eq:oscillation-phase} - \ref{eq:oscillation-intensity})
can be elicited by considering separately the behaviors of the oscillation phase $\phi$ and of the probability factor $\mathcal{F}$ of Eq. (\ref{eq:oscillation-intensity}).  

\begin{figure}[!htb]
\includegraphics[trim = 7mm 30mm 6mm 0mm, width=8cm]{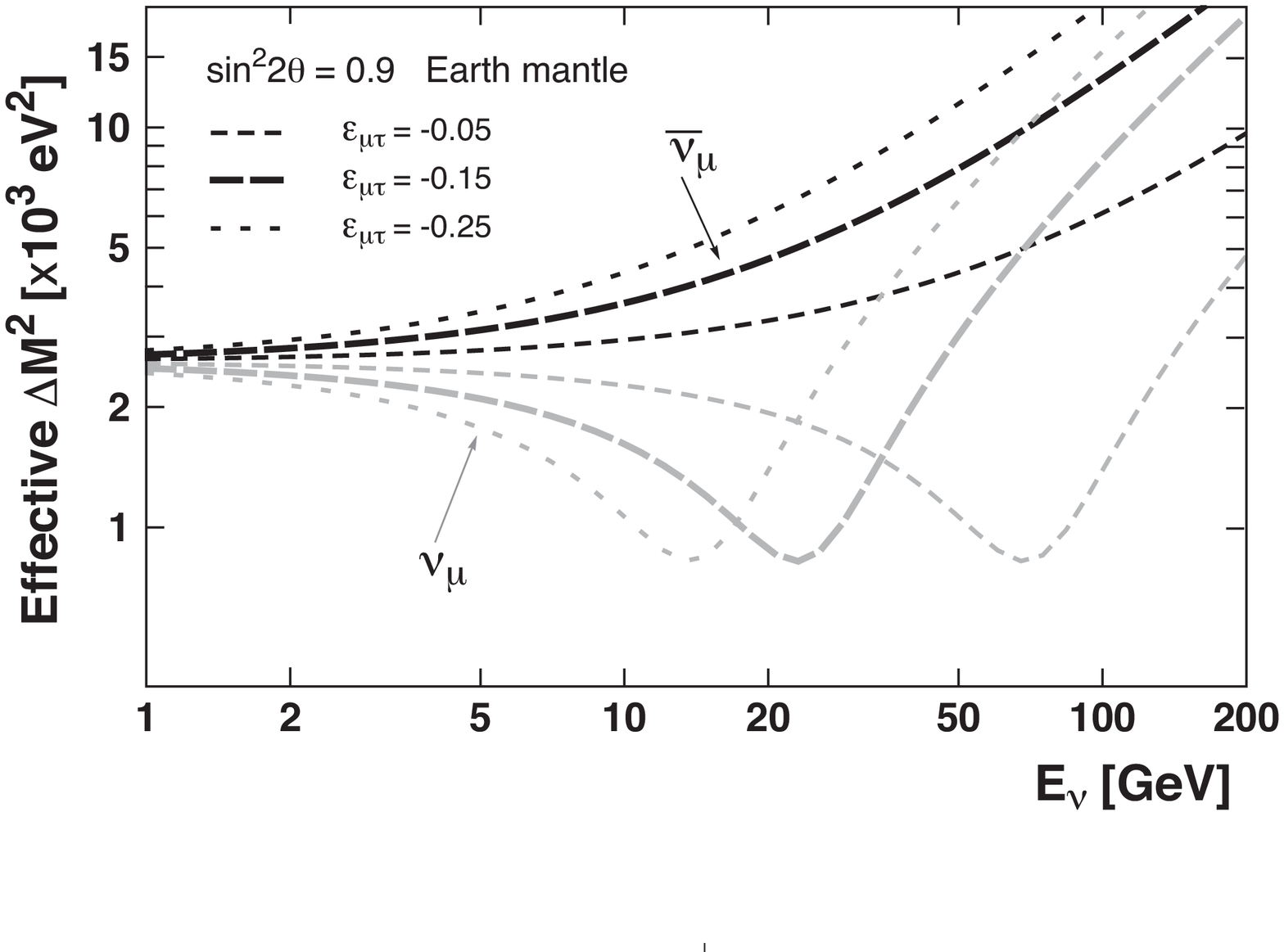}
\caption{
\small  The value of $\Delta M^{2}_{\mathrm{eff}}$ defined by Eq. (\ref{eq:effective-mass-relation}) for propagation in the mantle, as a function of neutrino energy.   
The NSI scenario suggests conventional analyses based upon vacuum oscillations may infer effective mass-squared differences that differ for $\numu$ and $\anumu$ oscillations and exhibit dependence upon $E_{\nu}$. 
The value $\Delta m^2_{32}=2.6 \times 10^{-3}\ \mathrm{eV}^2$ is assumed for the plotted curves.
}
\label{fig:DMeff-mantle}
\end{figure}

To examine the consequences of the matter effect for $\Delta m^2$ values extracted by analyses which assume conventional oscillations, we construct, as a toy model, the apparent or effective $\Delta M^2_{\mathrm{eff}}$. We equate the vacuum phase expression $\Delta M^2_{\mathrm{eff}} L/4 E_{\nu}$ to the matter effect oscillation phase $|\vec N| L$ of Eq. (\ref{eq:oscillation-phase}).
Then, 

\begin{eqnarray}  \label{eq:effective-mass-relation}  
\Delta M^{2}_{\mathrm{eff}} =  
\Delta m^{2}_{32} \Big\{ 1 ~\overline + ~ \sin2\theta \; \epsilon_{\mu \tau}\,|V_{e}| \left( 4 E_{\nu} /\Delta m^{2}_{32}\right) +  \nonumber  \\
 +   \left(\epsilon_{\mu \tau}V_{e}\right)^{2} \cdot \left( 4 E_{\nu} /\Delta m^{2}_{32}) \right)^{2}  \Big\}^{1/2} .\;\;\;\;
\end{eqnarray}   

\noindent          
For $\epsilon_{\mu \tau} \simeq -0.12$, the second term under the square root in Eq. (\ref{eq:effective-mass-relation}) is negative for $\numu$ and positive for $\anumu$.   
Equation (\ref{eq:effective-mass-relation}) implies separate curves for the apparent  $\Delta M_{\mathrm{eff}}^2$ as a function of $E_{\nu}$, for $\anumu$ and $\numu$ oscillations.   
The effect is illustrated in Fig. 4, which shows the $\Delta M_{\mathrm{eff}}^2$ curves for neutrino propagation through a terrestrial mantle of density 4.7 g/cm$^{3}$.  
A treatment of atmospheric event samples for which no event-by-event separation of antineutrinos and neutrinos is made, will likely converge on an `averaged' description
with $\Delta M_{\mathrm{eff}}^2 \simeq \Delta m_{32}^2$.   
On the other hand, in investigations in which $\anumu$ samples and $\numu$ samples are separately analyzed, two or more $\Delta M_{\mathrm{eff}}^2$ values may be inferred, depending upon the $E_{\nu}$ regimes which characterize the various event samples (e.g.  sub-GeV, multi-GeV contained, multi-GeV partially contained, upward-stopping muons, upward through-going muons, etc).

\begin{figure}[!htb]
\begin{center}
\includegraphics[trim = 10mm 10mm 2mm 0mm,width=8cm]{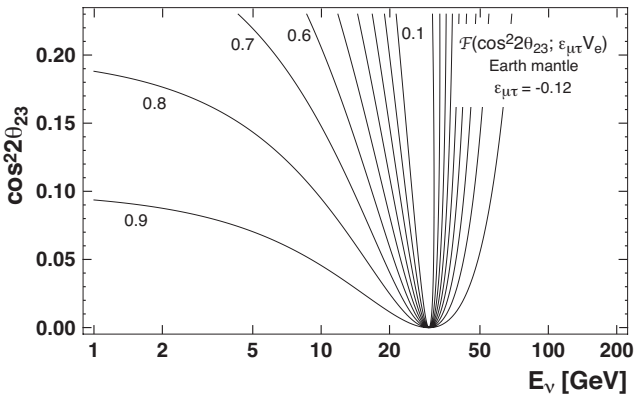}
\caption{
\small Nine contour lines showing the magnitude of the probability factor $\mathcal{F}$ for $\numu$ propagation in the mantle, Eq. (\ref{eq:oscillation-intensity}). The contours correspond to $\mathcal{F}=0.9, 0.8, ...,0.1$.
For atmospheric $\numu$'s with $E_\nu \approx 29~GeV$ and non-maximal $\theta_{23}$ mixing, the disappearance oscillation factor $\sin^2 \phi$ of Eq. (\ref{eq:oscillation-intensity}) is strongly damped. }
\label{fig:F-mantle}
\end{center}
\vspace{-0.7cm}
\end{figure}

For neutrino propagation in matter there arises an additional complication due to the probability factor $\mathcal{F}$ of Eq. (\ref{eq:oscillation-intensity}).   
For $\theta_{23}$ mixing which differs from maximal, the $\mathcal{F}$ factor introduces damping of the oscillation $\sin^{2}\phi$; 
the damping increases as $\cos^{2}2\theta_{23}$ is moved away from 0.  
Due to the sign difference in Eq. (\ref{eq:oscillation-intensity}), the damping is small for $\anumu$ propagation.  
However, for $\numu$ propagation at energies for which the negative term within the denominator of Eq. (\ref{eq:oscillation-intensity}) competes with the other two terms which are positive, the alteration of $\numu \rightarrow \numu$ survival probability can be dramatic.  
Figure~\ref{fig:F-mantle} illustrates the oscillation cloaking capability of the $\mathcal{F}$ factor.  
For $\numu$ propagation in the terrestrial mantle, oscillations of $\numu$ with $E_{\nu} \approx 29\;\mathrm{GeV}$ will be strongly damped with deviations from maximal mixing.   
This effect was identified and discussed previously in Ref.~\cite{OPERA} whose authors designated it as an ``anti-resonance".

The pivotal middle term within the denominator of Eq. (\ref{eq:oscillation-intensity}) that gives rise to different damping behavior of $\mathcal{F}$ for $\numu$ and $\anumu$ cases is proportional to $(\epsilon_{\mu\tau}V_{e} ) \times (\Delta m_{32}^2)^{-1}$.
We note that for the inverted hierarchy scenario which also fits the MINOS data, namely $(\Delta m_{32}^2, \epsilon_{\mu\tau}) = (-2.56 \times 10^{-3} \mathrm{eV}^2, +0.12)$, the anti-resonance remains to affect the $\numu$ propagation, and so the factor $\epsilon_{\mu\tau}V_{e}$ does not provide discrimination of the neutrino mass hierarchy.

\vspace{-10pt}
\subsection{Atmospheric neutrino experiments} \vspace{-9pt}
The differences between $\anumu$ and $\numu$ propagation in terrestrial mantle indicated by Figs.~\ref{fig:DMeff-mantle} and~\ref{fig:F-mantle}, may mislead atmospheric neutrino analyses which use events having good resolution for $L/E_{\nu}$ but fit combined samples of $(\numu + \anumu)$ with vacuum $\numu \rightarrow \nutau$ oscillation phenomenology.   
It is intriguing in this regard that the negative-log-likelihood contour for $\Delta m^2$ versus $\sin^{2}2\theta$ reported by the Soudan 2 collaboration does not exhibit a smoothly falling surface which converges to a single minimum.   
Rather, as shown in Fig. 9 of the experiment's final publication ~\cite{Soudan2}, the contour exhibits two local minima in the regime near maximal mixing;  
one minimum lies above the current global-average $\Delta m_{32}^2$ at $5.2 \times 10^{-3}~\mathrm{eV}^{2}$, while the other falls below, at $1.7 \times 10^{-3}\;\mathrm{eV}^{2}$. 
Similar behavior of a likelihood contour at maximal mixing was reported in the first MINOS analysis of 107 ($\numu + \anumu$) contained-vertex atmospheric events (see Fig. 14 of  \cite{MINOS-CVE}).  

Clearly of interest are atmospheric event samples wherein $\anumu$ and $\numu$ are distinguished, e.g. via magnetic tracking.   
The MINOS experiment has demonstrated such a capability with atmospheric neutrinos~\cite{MINOS-muons}.   
An analysis of a 24.6 kiloton-year exposure of the MINOS detector to atmospheric neutrinos is in progress~\cite{Tricia}. 

\vspace{-10pt}
\section{Discussion} \vspace{-9pt}

In view of the possible existence of an NSI matter effect $\epsilon_{\mu \tau}$ which may be large enough to distinguish $\anumu$
from $\numu$ disappearance oscillations in current-generation long baseline experiments, certain near-term developments are
worthy of note:
~${\it (i)}$  
As additional long baseline $\anumu$ data from MINOS becomes available, more direct determination of limits or else  
measurement of the sign and magnitude of $\epsilon_{\mu \tau}$ will become feasible.  
Currently, the MINOS collaboration is taking $\numu$ data with FHC running; 
plans are underway to enhance the $\anumu$ sample with more RHC running. ~${\it (ii)}$ 
Analysis of atmospheric neutrinos in which $\anumu$'s are examined separately from $\numu$'s using $\numu \rightarrow \nutau$ vacuum oscillation phenomenology, may obtain $\Delta \overline{m}^2$ which are larger than $\Delta m^2$.  
If event samples representing different energy regimes are separately analyzed, inferences of multiple  $\Delta ^{\scriptscriptstyle{(}}\overline m^{{\scriptscriptstyle{)}} 2}$ values are possible, along the lines indicated by Fig. 4.  
~${\it (iii)}$ 
An anti-resonance effect is predicted for high energy $\numu$ oscillations in matter for the case of non-maximal $\theta_{23}$ mixing.  
Its elucidation would be difficult for propagation baselines confined to the Earth's crust, however the effect might be accessible via study of $\numu$ propagation through the mantle.

The $\epsilon_{\mu \tau}$ matter effect scenario of this work differs from the two other proposals which were motivated by the early MINOS $\anumu$ results based upon FHC running.  
In CPT-violation hypotheses considered in ~\cite{Barenboim}, $\Delta \overline{m}_{32}^2$ and $\overline{\theta}_{23}$ are to be distinguished from $\Delta m_{32}^2$ and $\theta_{23}$, whereas in our view such distinctions are a mirage arising from application of incomplete phenomenology.  
Our outlook is more nearly akin to the ``apparent CPT violation" proposal of Ref.~\cite{Nelson}.   
However in the latter proposal, a light sterile neutrino conspires with a $B-L$ interaction to generate a difference between $\anumu$ and $\numu$ disappearance.   
A direct manifestation of mixing of active neutrinos to a sterile neutrino is that the ratio $R$ of observed to expected neutral current rate at a far detector site is predicted to be less than 1.0 for either of $\numu$ or $\anumu$ exposures \cite{MINOS-NC}.   
In contrast, our NSI matter effect gives rise to probability oscillations among active neutrino flavors, but with no net flavor loss, hence $R$ measurements are expected to yield 1.0.   

Other proposals for new physics to distinguish $\anumu$ from $\numu$ oscillations \cite{Heeck, Choudhury}, and remarks
on using NSI for this purpose \cite{Akhmedov}, have appeared in recent preprints.   Implications for
accelerator neutrino experiments arising from neutral current NSI, and a possible role for charged current NSI 
of the type $\nu_\tau + N \rightarrow \mu + X$, are discussed in \cite{Parke}.    The phenomenological exposition presented here, as well as those to be found in all of  the other above-mentioned works, have one theme in common:  The data currently emerging from $\anumu$ exposures at various baselines deserves careful scrutiny for evidence of particle interactions which lie outside the purview of the Standard Model.

\vspace{+7pt}
\section{Acknowledgments} \vspace{-8pt}

This study was conducted independently of the analysis deliberations of the MINOS Collaboration.   
In particular, all information used in fitting is readily available as reported by the Collaboration in Ref. \cite{Tricia}.   
Nevertheless it is the case that the neutrino oscillation observations of MINOS provided a strong motivation for this investigation.    
We are indebted to our MINOS colleagues for their dedicated work over many years which has realized the measurements recently reported.    
Special thanks is owed to William Oliver for critical readings and helpful discussions concerning the physics of matter effects.  
This work was supported by the United States Department of Energy under grant DE-FG02-92ER40702.

\end{document}